\documentclass[twocolumn,showpacs,preprintnumbers,amsmath,amssymb,prb]{revtex4}

\usepackage{graphicx}
\usepackage{dcolumn}
\usepackage{bm}

\begin{document}

\input epsf.sty

\widetext

\title
{
Structural effect on the static spin and charge correlations in La$_{1.875}$Ba$_{0.125-x}$Sr$_{x}$CuO$_{4}$ 
}


\author{M. Fujita}
\email{fujita@scl.kyoto-u.ac.jp}
 \author{H. Goka}%
 \author{K. Yamada}%
\affiliation{%
Institute for Chemical Research, Kyoto University, Uji 610-0011, Japan
}%

\author{M. Matsuda}

\affiliation{%
Advanced Science Research Center, Japan Atomic Energy Research Institute, Tokai 319-1195, Japan
}


\date{\today}



\begin{abstract}
We report the results of elastic neutron scattering measurements performed on 1/8-hole doped La$_{1.875}$Ba$_{0.125-x}$Sr$_{x}$CuO$_{4}$ single crystals with {\it x}=0.05, 0.06, 0.075 and 0.085. In the low-temperature less-orthorhombic (LTLO, {\it Pccn} symmetry) phase, the charge-density-wave (CDW) and spin-density-wave (SDW) wavevectors were found to tilt in a low-symmetric direction with one-dimensional anisotropy in the CuO$_{2}$ plane, while they were aligned along the high-symmetry axis in the low-temperature tetragonal (LTT, {\it P}4$_2$/{\it ncm} symmetry) phase. The coincident direction of two wavevectors suggests a close relation between CDW and SDW orders. The SDW wavevector systematically deviates from the Cu-O bond direction in the LTLO phase upon Sr substitution and the tilt angle in the LTLO phase is smaller than that in the low-temperature orthorhombic phase (LTO, {\it B}{\it mab} symmetry) with comparable in-plane orthorhombic distortion. These results demonstrate a correlation between the corrugated pattern of CuO$_{2}$ plane and the deviations. 
\end{abstract}



\maketitle

\section{Introduction}
Since the discovery of SDW and CDW orders in a lamellar copper-oxide ~\cite{Tranquada95}, much interest has been focused on the role of stripe correlations for high-{\it T}$_{c}$ superconductivity.~\cite{Emery97,Castellani97,Vojta99} 
A systematic study of the tetragonal La$_{2-x}$Nd$_{0.4}$Sr$_{x}$CuO$_{4}$ (LNSCO) system revealed a competitive relationship between the stripe order and superconductivity.~\cite{Ichikawa00,Tranquada97} 
However, for the orthorhombic systems La$_{2}$CuO$_{4+y}$ (LCO)~\cite{YoungLee99} and La$_{1.88}$Sr$_{0.12}$CuO$_{4}$ (LSCO)~\cite{Kimura99}, the SDW order was found to coexist with superconductivity and no well-defined incommensurate (IC) peak from the CDW order was observed. 
These results suggest a relation between the stripe order, superconductivity and a crystal structure. 
Recent neutron scattering measurements on a 1/8-hole doped La$_{1.875}$Ba$_{0.125-x}$Sr$_{x}$CuO$_{4}$ (LBSCO) system~\cite{Fujita_cond-mat} clearly demonstrate a distinct structural effect on SDW and CDW orders; 
the CDW order is stabilized in the LTT and LTLO phases and dramatically degraded towards the LTO phase, while peak-intensity from the SDW order remains even in the LTO phase. 
Moreover, suppression of the superconductivity is coupled with the CDW order.~\cite{Fujita_cond-mat} 

High-resolution neutron scattering measurements revealed a further difference between SDW peak-positions in the tetragonal LNSCO system and those in orthorhombic LCO and LSCO systems. 
Observed quartet SDW peaks in the former system form a fourfold square-shaped arrangement around ($\pi$,$\pi$)~\cite{Tranquada_private}, while those in the latter two systems are located at corners of rectangular with twofold symmetry~\cite{YoungLee99,Kimura00} as illustrated in Figs. 1(a) and (b). 
This peak-shift to low-symmetric positions in the ({\it h} {\it k} 0) plane, the {\it Y}-{\it shift}, which was discovered by Lee and co-workers\cite{YoungLee99}, indicates a deviation of the SDW wavevector from the underlying Cu-O bond direction with one-dimensional anisotropy in the CuO$_{2}$ plane. 
On the other hand, all observed IC peaks from the CDW order in the LNSCO system are located at high-symmetric positions.~\cite{Tranquada95,Ichikawa00,Tranquada97} 
Thus, further clarification of the relationship between peak-shifts and the crystal structure as well as greater understanding of the {\it Y}-shift of CDW peaks is needed in order to study the origins of SDW and CDW orders. 
Especially in the stripe model, the {\it Y}-shift of CDW peaks should be 
\linebreak
\begin{figure}[t]
\centerline{\epsfxsize=3.1in\epsfbox{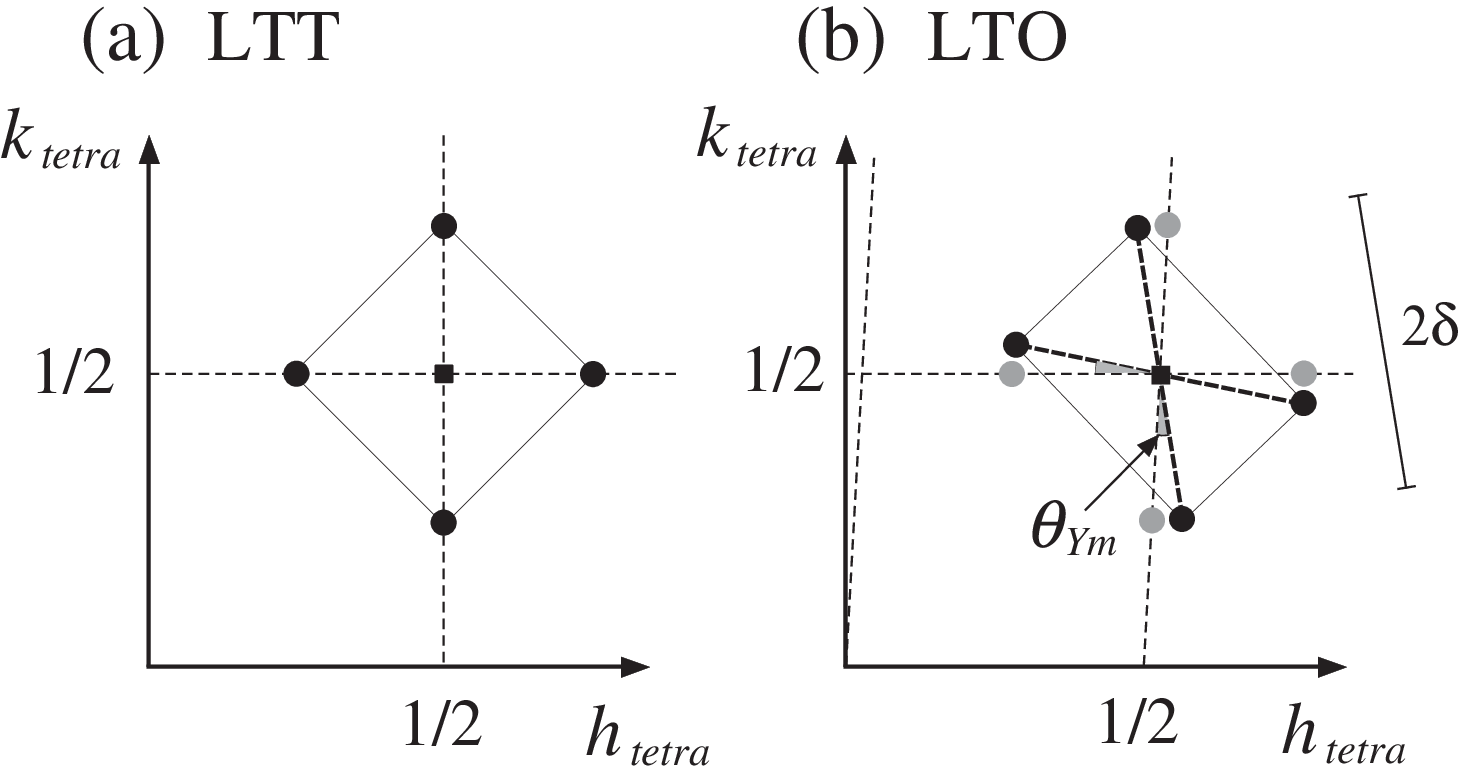}}
\caption
{
Observed positions of SDW peaks (closed circles) around the magnetic zone-center (closed squares) in the (a) LTT~\cite{Tranquada_private,Fujita01} and (b) LTO~\cite{YoungLee99,Kimura99} phases. Thin dashed lines correspond to the high-symmetry axis in the reciprocal lattice. Gray circles in Fig. 1(b) represent the peak-position without the {\it Y}-shift. 
}
\end{figure}
%
\noindent
observed in the orthorhombic phase because a tilt of the charge stripe is responsible for this peak-shift.~\cite{Bosch01} 

To address these issues, we have performed comprehensive elastic neutron scattering measurements on 1/8 hole-doped LBSCO single crystals. 
The {\it Y}-shift of SDW and CDW peaks in the LTLO phase have been observed for the first time. 
The tilt directions of the two wavevectors are same, suggesting a close relation between SDW and CDW orders. 
Furthermore, the tilt angle of the SDW wavevector changes continuously through the LTLO phase, varying with the Ba/Sr ratio while maintaining a constant carrier concentration of 1/8. 
Thus, the {\it Y}-shift is affected by the crystal structure. 

The format of this paper is as follows. Sample preparation and experimental details are described in Section II. The results of elastic neutron scattering measurements are introduced in Section III. Finally, in Section IV we discuss the relationship between the {\it Y}-shift and crystal structure, and elaborate on possible origins of this shift. 

\section{EXPERIMENTAL DETAILS}

Single crystals of LBSCO ({\it x}=0.05, 0.06, 0.075 and 0.085) are grown by a standard traveling-solvent floating-zone method. 
The shape of as-grown crystals is columnar with typical dimensions of $\sim$6 mm in diameter and $\sim$100 mm in length. 
For neutron scattering measurements, crystal rods near the final part of the growth were cut into $\sim$30 mm long pieces. 
To minimize oxygen deficiencies, crystals are then annealed under oxygen gas flow at 900 $^{\circ}$C for 50 h, cooled to 500 $^{\circ}$C at a rate of 10 $^{\circ}$C/h, annealed at 500 $^{\circ}$C for 50 h, and finally furnace-cooled to room temperature. 
Superconducting shielding signals are measured with a superconducting quantum interference device magnetometer in order to determine superconducting transition temperatures {\it T}$_{c}$. 
Evaluated {\it T}$_{c}$(onset)$^{\prime}$s are 10.0 K, 11.5 K, 14.0 K and 32.0 K for {\it x}=0.05, 0.06, 0.075 and 0.085 samples~\cite{Goka_unpublish,Fujita01}, consistent with results for polycrystalline samples.~\cite{Maeno91} 

In La-214 systems, coherent tilt of CuO$_{6}$ octahedron leads successive structural phase transition. 
Crystal structure of the present {\it x}=0.05 sample is LTT and that of the {\it x}=0.06, 0.075 and 0.085 samples is LTLO at low temperature and the transition temperatures between LTO and LTT/LTLO ({\it T}$_{d2}$) are 37K, 35K, 32K and 30K, respectively. 
The in-plane lattice constants of tetragonal {\it x}=0.05 and orthorhombic {\it x}=0.085 samples, for instance, are {\it a}$_{tetra}$={\it b}$_{tetra}$=3.787 $\AA$ and {\it a}$_{ortho}$=5.327 $\AA$ and {\it b}$_{ortho}$=5.361 $\AA$ below 6K. 
Note that the orthorhombic lattice constant is approximately $\sqrt{2}$ times of tetragonal one. 
The orthorhombic phase typically includes four possible domains, however, small number of domains are suitable for the quantitative analysis of IC peaks. 
Fortunately, all of present crystals with the LTLO phase have two domains and the IC peak-geometry is relatively simple, that is, the superposition of those for two different single domains. 
(IC peak-geometry expected from stripe order for one of single-domain with the {\it Y}-shift is shown in Fig. 2(a).) 
In Fig. 2(b), superposed CDW peak positions are shown by circles. Closed and open symbols indicates signals from different domains. 
A part of peak-positions is clarified by the present study as the result is shown in the next section. 
Gray circles in the inset of Fig. 2(b) represent CDW peak-positions without the {\it Y}-shift. 
Thus, when the CDW wavevector is aligned along a high-symmetry axis, i.e. the Cu-O bond direction, wider peak-splitting should be seen in the transverse-scan (thick arrows) at larger {\it k}, while the peak-splitting narrows (widens) at larger (smaller) {\it k} with increasing values of the {\it Y}-shift, $\theta$$_{Ych}$.

Neutron scattering measurements were performed using TOPAN, TAS-1 and HER triple-axis spectrometers installed in the JRR-3M reactor at the Japan Atomic Energy Research Institute (JAERI). 
\linebreak
\begin{figure}[t]
\centerline{\epsfxsize=2.2in\epsfbox{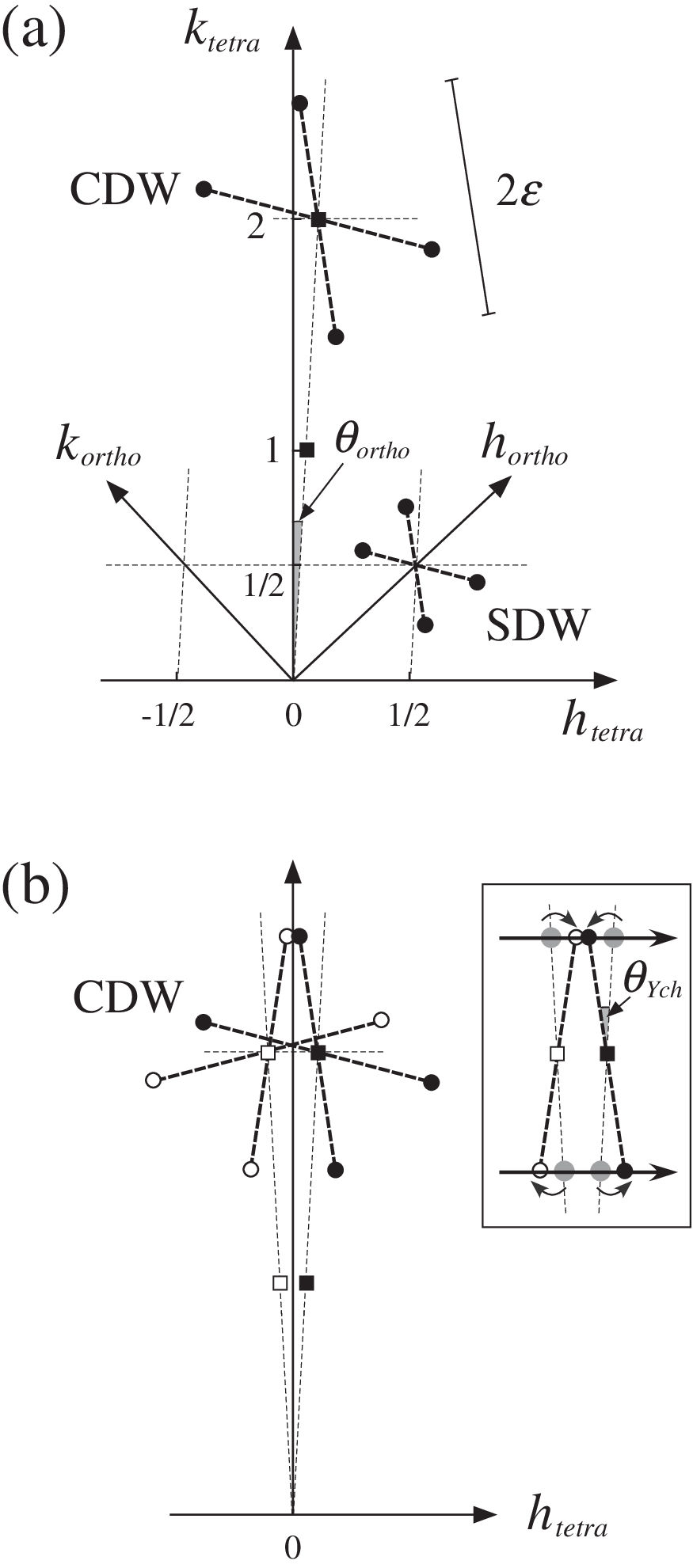}}
\caption
{
eak-geometry of (a) SDW and CDW peaks in the ({\it h} {\it k} 0) scattering plane for the single-domain orthorhombic sample and (b) that of CDW peaks for the two-domain sample. Nuclear Bragg and supperlattice peaks are indicated by squares and circles, respectively. Open and closed symbols represent signals from different domains. Gray circles in the inset of Fig. 2(b) are a number of CDW peak-positions without the {\it Y}-shift. 
}
\end{figure}
\noindent
We measured IC peaks from the CDW order at the thermal neutron TOPAN and TAS-1 spectrometers with horizontal collimator sequences of 30$^{\prime}$-30$^{\prime}$-{\it S}-30$^{\prime}$-150$^{\prime}$ and 80$^{\prime}$-40$^{\prime}$-{\it S}-40$^{\prime}$-150$^{\prime}$, respectively, where {\it S} denotes the sample position. 
The incident and final neutron energies were fixed at 14.7 meV for TOPAN and 13.7 meV for TAS-1 using the (0 0 2) 
reflection from pyrolytic graphite crystals. 
Most experiments for investigating the SDW peak-geometry were performed using the cold neutron HER spectrometer with high experimental resolution. 
In this case, the fixed neutron energy and horizontal collimation were 5.0 meV and 32$^{\prime}$-100$^{\prime}$-{\it S}-80$^{\prime}$-80$^{\prime}$. 
In order to eliminate the higher-order reflected beams, pyrolytic graphite and beryllium filters were used at thermal and cold neutron spectrometers, respectively. 
Crystals were mounted in the ({\it h} {\it k} 0) zone and cooled using a $^{4}$He-closed cycle refrigerator, or a top-loading liquid-He cryostat. 
In this paper, crystallographic indices are denoted by the tetragonal {\it I}4/{\it mmm} notation in order to express scans done for investigating {\it Y}-shift easily, even though the crystals is orthorhombic at low temperature. 

\section{results}

\subsection{Charge-density-wave order}
CDW peak-profiles measured along the transverse 
\linebreak
\begin{figure}[b]
\centerline{\epsfxsize=2.3in\epsfbox{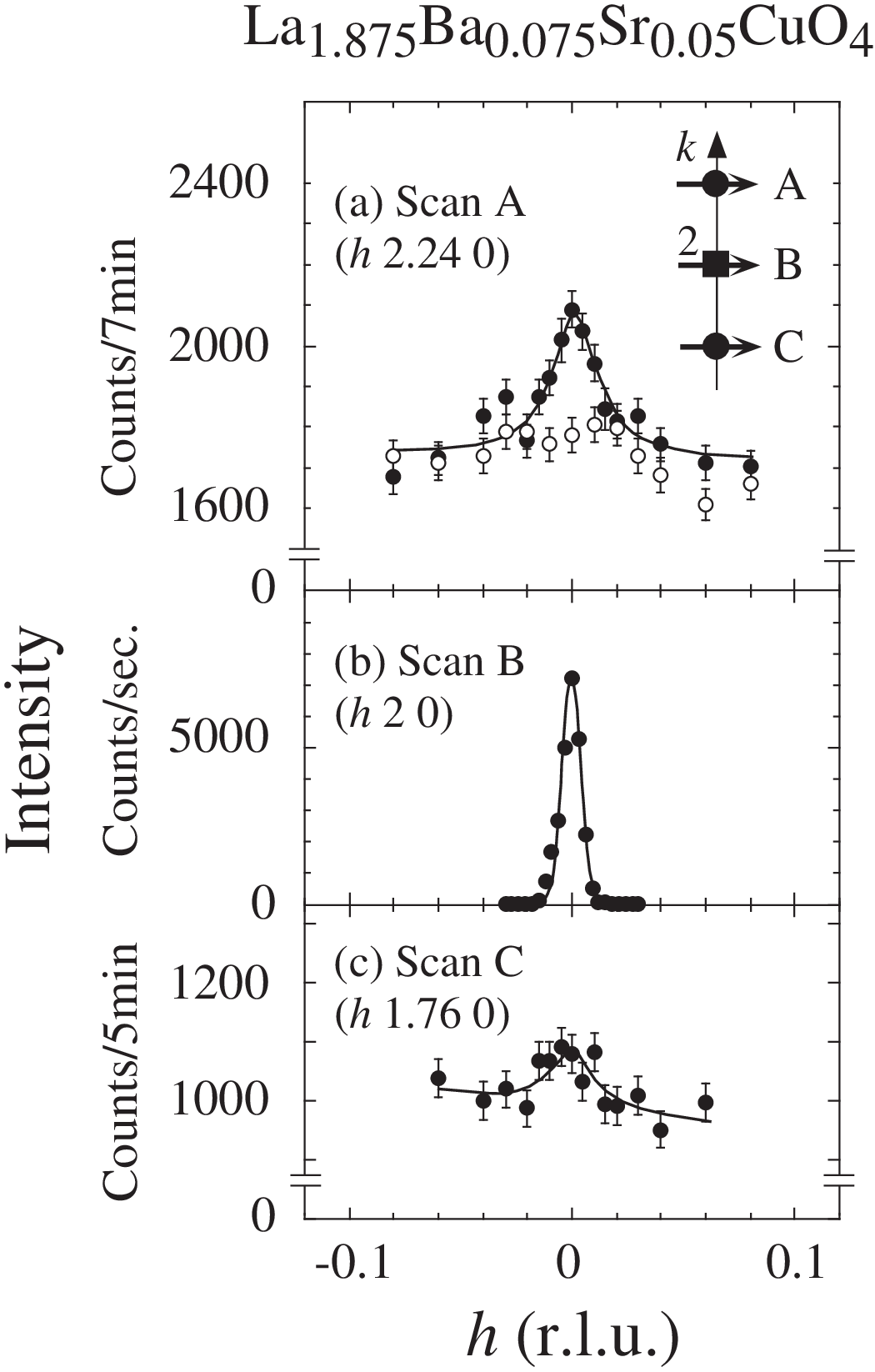}}
\caption
{
Profiles of (a), (c) CDW peaks and the (b) (0 2 0) nuclear Bragg peak for the tetragonal La$_{1.875}$Ba$_{0.075}$Sr$_{0.05}$CuO$_{4}$ sample measured using the TOPAN spectrometer. Solid and open circles represent data taken at 1.4 K ($<$ {\it T}$_{d2}$) and 45 K ($>$ {\it T}$_{d2}$). Scan trajectories and CDW (Bragg) peak-positions are shown by arrows and circles (squares) in the inset of (a), respectively. 
}
\end{figure}
\noindent
direction for the LTT phase of La$_{1.875}$Ba$_{0.075}$Sr$_{0.05}$CuO$_{4}$ are shown in Figs. 3(a) and (c). 
Solid lines in the figures are results fitted with a single Lorentzian function by convoluting the experimental resolution. 
The centers of each peak are found to be located at {\it h}=0. 
As shown in Fig. 3(b), the nuclear Bragg peak is also positioned at {\it h}=0.
These results demonstrate the parallel/perpendicular CDW wavevector to the Cu-O bond direction as seen in the tetragonal phase of La$_{1.48}$Nd$_{0.4}$Sr$_{0.12}$CuO$_{4}$.~\cite{Tranquada95,Ichikawa00,Tranquada97} 
The resolution-corrected peak-width, $\kappa$$_{ch}$ of 0.009$\pm$0.001 $\AA$$^{-1}$ in half width at half maximum (HWHM) corresponding to a correlation length, $\xi$$_{ch}$ of $\sim$110 $\AA$ for lattice distortion and is consistent with the value obtained from a recent high energy X-ray diffraction measurement.\cite{Kimura_unpub} 
We mention here that the width of the (0 2 0) Bragg peak scanned along the transverse direction is resolution limited. Thus, the quality of a crystal with a mosaic spread of $\leq$0.1$^{\circ}$ (HWHM) is quite 
good for the present measurements.

Figure 4. shows the CDW peaks in the LTLO phase of La$_{1.875}$Ba$_{0.05}$Sr$_{0.075}$CuO$_{4}$ scanned along identical trajectories as the previously mentioned {\it x}=0.05 sample. 
\linebreak
\begin{figure}[b]
\centerline{\epsfxsize=2.3in\epsfbox{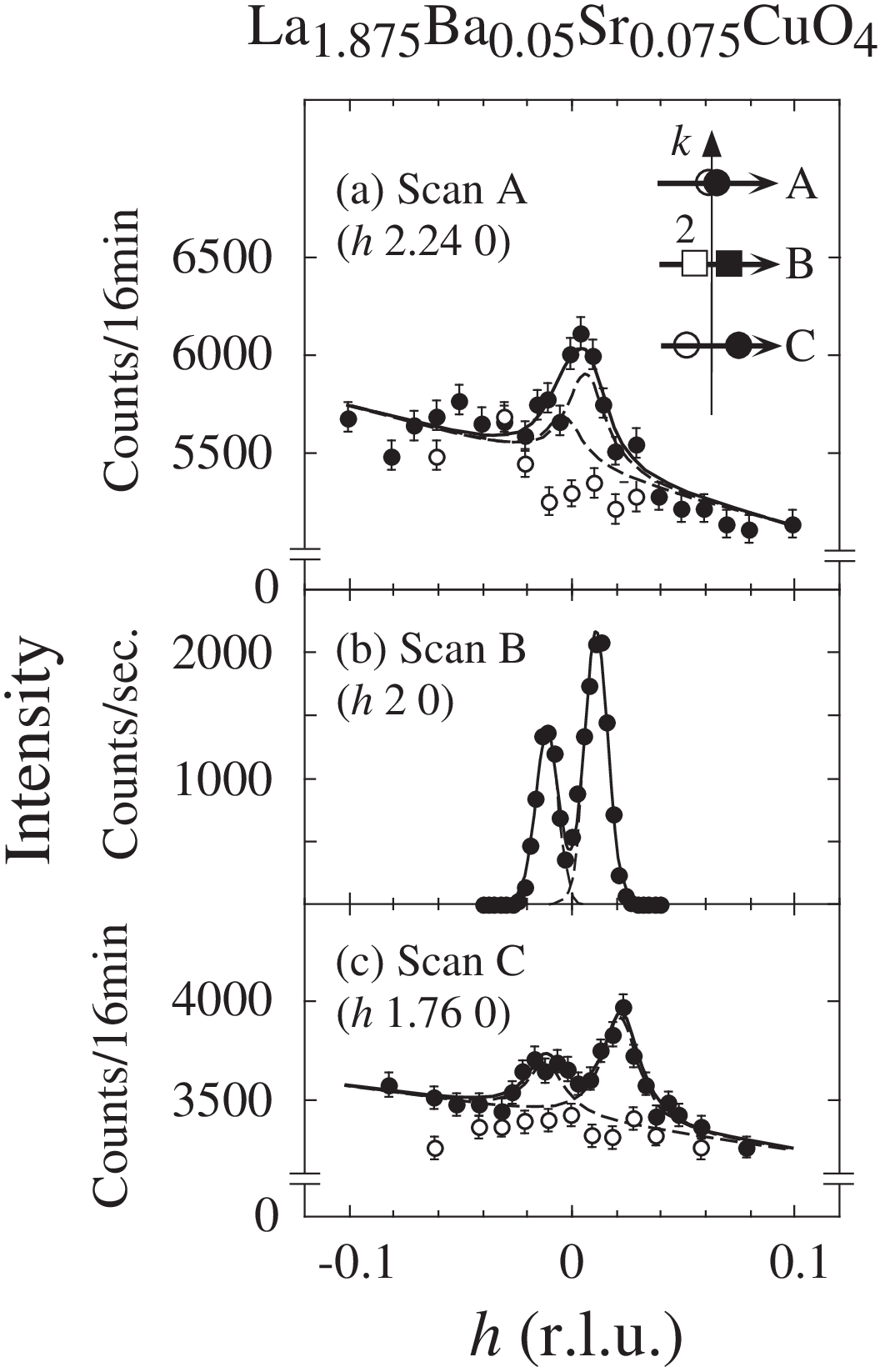}}
\caption
{
Profiles of (a), (c) CDW peaks and (b) nuclear Bragg peaks around (0 2 0) for the orthorhombic La$_{1.875}$Ba$_{0.05}$Sr$_{0.075}$CuO$_{4}$ sample measured using the TAS-1 spectrometer. Solid and open circles represent the data taken at 5 K ($<$ {\it T}$_{d2}$) and 40 K ($>$ {\it T}$_{d2}$). Scan trajectories and CDW (Bragg) peak-positions are shown by arrows and circles (squares) in the inset of (a), respectively. 
}
\end{figure}
\noindent
As seen in Figs. 4(a) and (c), in scans along the {\it h}-direction a single peak was observed at {\it k}=2.24, while peaks are split at k=1.76, indicating the {\it Y}-shift of CDW peaks as explained in Sec. II. 
For quantitative analysis,we have assumed two pairs of CDW peaks centered at each (0 2 0) Bragg peak (Fig. 4(b)), consistent with the stripe order in a two-domain sample. (See inset of Fig. 2(b).) 
Then we fitted the profiles scanned along A and C to the following Lorentzian function: 
\begin{displaymath}
   \frac{1} {\{{h}+\eta_{1,2}\}^2+\kappa_{ch}^2}+\frac{1.6} {\{{h}-\eta_{1,2}\}^2+\kappa_{ch}^2}  
\hspace{5mm}
\end{displaymath}
where $\eta$$_{1}$ and $\eta$$_{2}$ denote peak-positions in the {\it h}-direction at {\it k}=1.76 and 2.24, respectively.
Since the intensities of two Bragg peaks, which correspond to relative population of domains, are unbalanced with a ratio of 1:1.6, a proportional factor of 1.6 is added in the second term of above expression. 
The solid lines in Figs. 4(a) and (c) are fitted results while dashed lines represent individual contributions. 
The evaluated values for $\eta$$_{1}$ and $\eta$$_{2}$ of 0.017$\pm$0.001 and 0.005$\pm$0.001(r.l.u.), respectively, lead to a value of 1.7$\pm$0.3$^{\circ}$ for $\theta$$_{Ych}$, which is defined in Fig. 2(b) as being equal to the tilt angle of the CDW wavevector from the high-symmetry axis. 
$\kappa$$_{ch}$ is found to be 0.010$\pm$0.001 $\AA$$^{-1}$, similar to the result of {\it x}=0.05. 
Only a single pair of the expected four CDW peaks from each domain was observed within experimental error, as the similar result was reported for the LNSCO system.\cite{Tranquada96} 
We note that the integrated intensities for ({\it h} 1.76 0) and ({\it h} 2.24 0) CDW peaks are same contrasitve to the case of LNSCO system as expected from simple stripe model\cite{Tranquada96}, possibly due to the effect from the out-of-plane atomic displacements. 

\begin{figure}[b]
\centerline{\epsfxsize=2.2in\epsfbox{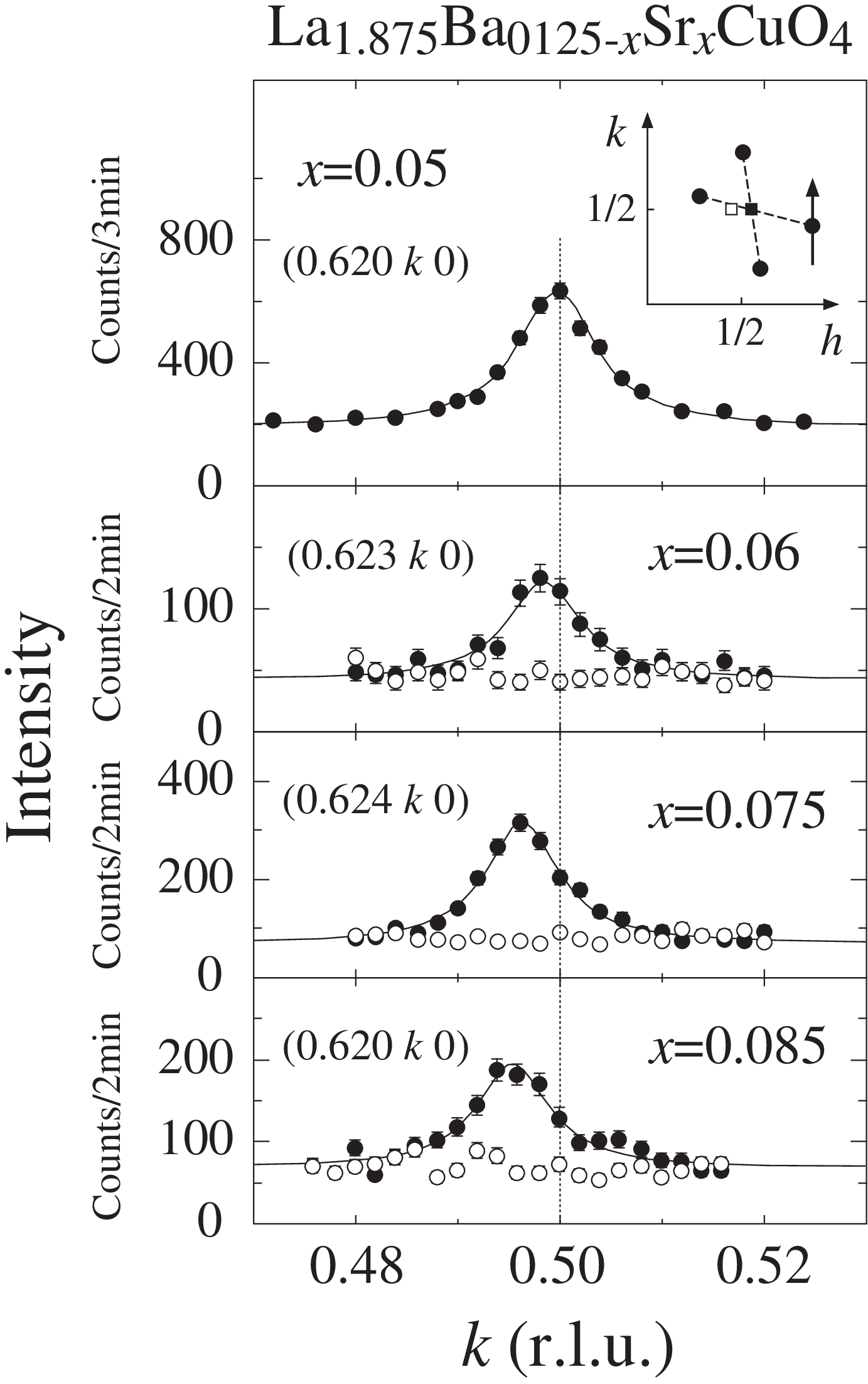}}
\caption
{
SDW peak-profiles for {\it x}=0.05, 0.06, 0.075, and 0.085 samples measured using the HER spectrometer. Solid and open circles represent data taken at 6 K ($<$ {\it T}$_{d2}$$^{\prime}$s) and 45 K ($>$ {\it T}$_{d2}$$^{\prime}$s). The scan geometry is depicted in the inset of the upper panel. 
}
\end{figure}

\subsection{Spin-density-wave order}

A tilt of the SDW wavevector was also found in the LTLO phase. 
In Fig. 5 we present a series of IC peaks from the SDW order taken along the scan trajectory shown in the inset. 
In the tetragonal {\it x}=0.05 sample peak is located at {\it h}=0.5, however, in orthorhombic samples, the peak-center anisotropically shifts from {\it h}=0.5 and the magnitude increases with increasing Sr concentration, {\it x}. 
Observed peak-geometries in the LTLO phase are identical to those in the LTO phase.~\cite{YoungLee99,Kimura00} 
The SDW peak-width for each sample is resolution limited, corresponding to the magnetic correlation length, $\xi$$_{m}$ $\geq$200 $\AA$.
Note that the systematic peak-shift in the region comparable with the peak-width contradicts the macroscopic phase separation of LTT and LTLO/LTO in the present single crystals. 

In Fig. 6 (a), the SDW wavevector tilt angle, $\theta$$_{Ym}$, as a function of Sr concentration is shown together with the CDW angle, $\theta$$_{Ych}$. 
Similar to the case of $\theta$$_{Ych}$, $\theta$$_{Ym}$ is equal to the shift of magnetic IC peaks from 
the high-symmetric direction in an angle unit in the reciprocal lattice (see Fig. 1). 
Thus, we evaluated $\theta$$_{Ym}$ from the positions of paired IC peaks. 
\linebreak
\begin{figure}[b]
\centerline{\epsfxsize=2.7in\epsfbox{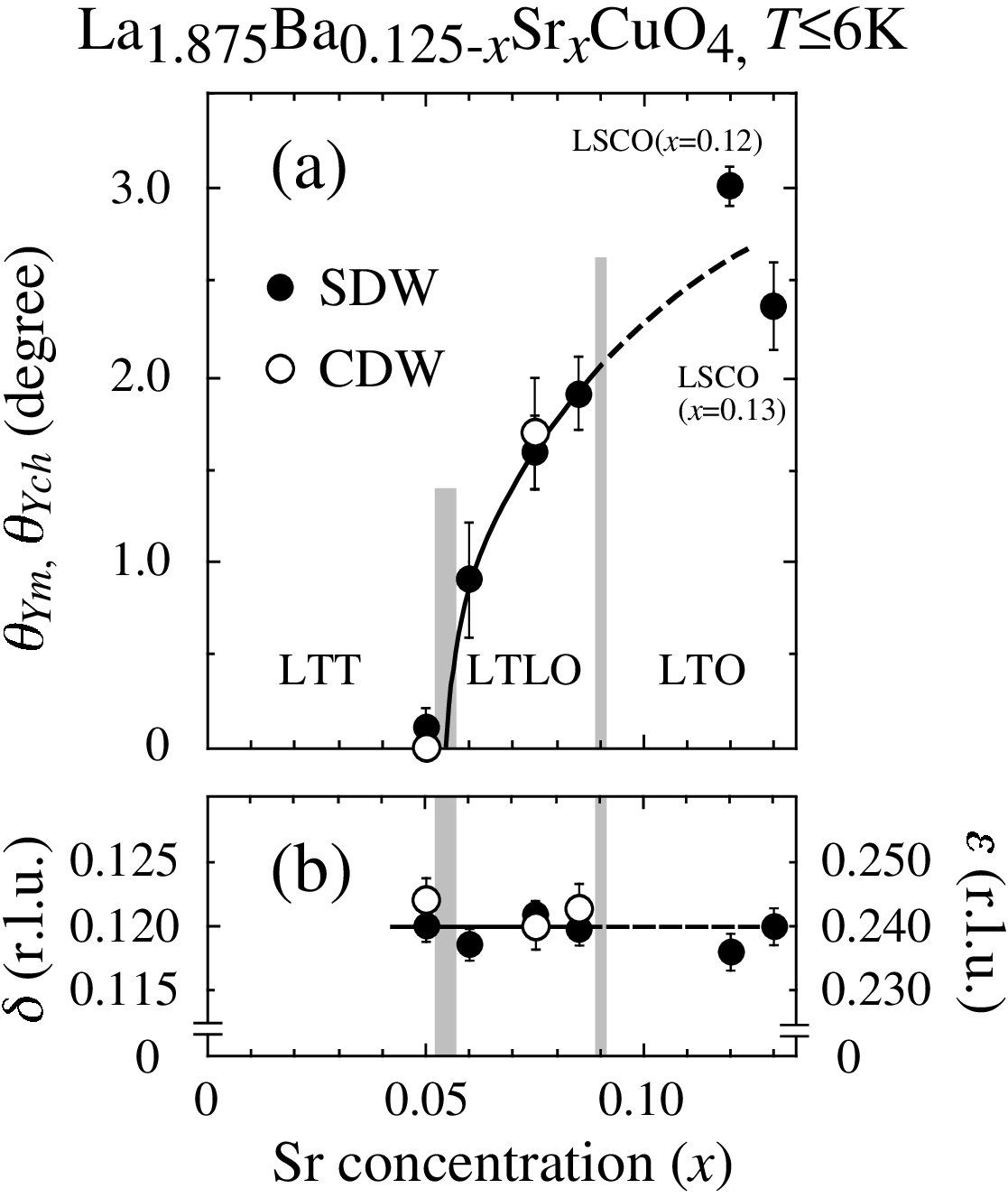}}
\caption
{
Sr concentration dependence of (a) the tilt angles, $\theta$$_{Ym}$ and $\theta$$_{Ych}$, and (b) the incommensurabilities, $\delta$ and $\epsilon$ for SDW (solid circles) and CDW (open circles) orders. Previously reported $\theta$$_{Ym}$ values for La$_{1.88}$Sr$_{0.12}$CuO$_{4}$~\cite{Kimura00} and La$_{1.87}$Sr$_{0.13}$CuO$_{4}$~\cite{Matsushita99} are also plotted as references. Shaded bars represent the structural phase boundary.~\cite{Fujita_cond-mat} The solid and dashed lines are guides to the eye. 
}
\end{figure}
\noindent
Considering previous results showing modification of the crystal structure with variation of the Ba/Sr ratio\cite{Fujita_cond-mat}, the {\it Y}-shift coincidentally appears with the LTLO phase and $\theta$$_{Ym}$ increases in the orthorhombic phase upon Sr substitution. 
Even more importantly, $\theta$$_{Ym}$ and $\theta$$_{ch}$ in both the LTT phase with {\it x}=0.05 and the LTLO phase with {\it x}=0.075 are equivalent, demonstrating a close relation between SDW and CDW orders and supporting the existence of stripe order in these phases. 
In contrast to {\it x}-dependent $\theta$$_{Ym}$ and $\theta$$_{ch}$, the incommensurabilities $\epsilon$ and $\delta$ for CDW and SDW orders, defined as the half distance between each IC peak (See Figs. 1(a) and 2(a).), are constant within the error (Fig. 6(b)) and satisfy the simple relation of $\epsilon$=2$\delta$ irrespective of the crystal structure. 
We mention that $\delta$ is close to total hole concentration. 
\linebreak
\begin{table}[b]
\caption{\label{tab:fonts} 
Tilting angle for the SDW peak $\theta$$_{Ym}$ and in-plane orthorhombic distortion $\theta$$_{ortho}$ of the LTT/LTLO phase of La$_{1.875}$Ba$_{0.125-x}$Sr$_{x}$CuO$_{4}$ (upper row) and the LTO phase of Zn-doped and free LSCO and the LCO (lower row) systems.
}
\begin{ruledtabular}
\begin{tabular}{p{42mm}p{12mm}cl}
{Sample} & {$\theta$$_{Ym}$} ($^{\circ}$) & {$\theta$$_{orhto}$ ($^{\circ}$)} & {Ref.}\\
\hline
{La$_{1.875}$Ba$_{0.125-x}$Sr$_{x}$CuO$_{4}$} \\
{\hspace{2mm} {\it x}=0.05} & {0.1$\pm$0.1} & {0} \\
{\hspace{2mm} {\it x}=0.06} & {0.9$\pm$0.4} & {0.27} \\
{\hspace{2mm} {\it x}=0.075} & {1.6$\pm$0.2} & {0.31} \\
{\hspace{2mm} {\it x}=0.085} & {1.9$\pm$0.2} & {0.37} \\
\hline
{La$_{1.79}$Sr$_{0.21}$Cu$_{0.99}$Zn$_{0.01}$O$_{4}$} & {2.1$\pm$0.3} & {$\leq$0.10} & {[18]}\\
{La$_{1.87}$Sr$_{0.13}$CuO$_{4}$} & {2.4$\pm$0.3} & {0.26} & {[19]} \\
{La$_{1.88}$Sr$_{0.12}$CuO$_{4}$} & {3.0$\pm$0.1} & {0.32} & {[11]} \\
{La$_{2}$CuO$_{4+y}$} & {3.3$\pm$0.1} & {0.36} & {[7]} \\
\end{tabular}
\end{ruledtabular}
\end{table}
%
\noindent

\section{discussion and conclusions}

One of the most important results of this study is that the tilt of the CDW wavevector from the high-symmetry axis is clearly observed in the LTLO phase, and this deviation is coincident with that of the SDW wavevector. 
Combining these results with the fact that the wavevectors are coincidentally aligned along the high-symmetry axis in the LTT phase, $\theta$$_{ch}$ possibly shows similar Sr concentration dependence with $\theta$$_{Ym}$. 

As shown in Table I, the in-plane orthorhombic lattice distortion in angle units, $\theta$$_{ortho}$, (defined in Fig. 2(a)) becomes large in the present LBSCO system with Sr substitution, indicating change in the tilt angle of the CuO$_{6}$ octahedron. 
$\theta$$_{Ym}$ increases approximately in proportion to ($\theta$$_{ortho}$)$^{2}$. 
These results demonstrate a close relationship between the {\it Y}-shift and crystal structure, as the total hole concentration is fixed. 
An analogous increase in $\theta$$_{Ym}$ for $\theta$$_{ortho}$ is seen in the LSCO and LCO systems~\cite{calc}, although in this case both the crystal structure and the hole concentration change. 
However, $\theta$$_{Ym}$ in the LTLO phase of LBSCO is smaller than that in the LTO phase of Zn-doped and free LSCO~\cite{Kimura00,Matsushita99,Kimura01} and LCO~\cite{YoungLee99} with a comparable in-plane distortion, in which no well-defined CDW peak is observed. 
This is true even when comparing values among samples with hole concentrations near 1/8. 
Thus, the tilt angle of SDW and CDW wavevectors correlates with the corrugated pattern of the CuO$_{2}$ plane or with the tilt angle of the CuO$_{6}$ octahedron rather than with the in-plane lattice distortion, although the deviation itself may be a common feature in the orthorhombic phase of La-214 systems.~\cite{Kimura_private} 
On the other hand, we reported previously that the stabilization of SDW and CDW orders in the LBSCO system is also related to the crystal structure, as both orders are well stabilized in the LTT phase.\cite{Fujita_cond-mat} 
\linebreak
\begin{figure}[t]
\centerline{\epsfxsize=2.55in\epsfbox{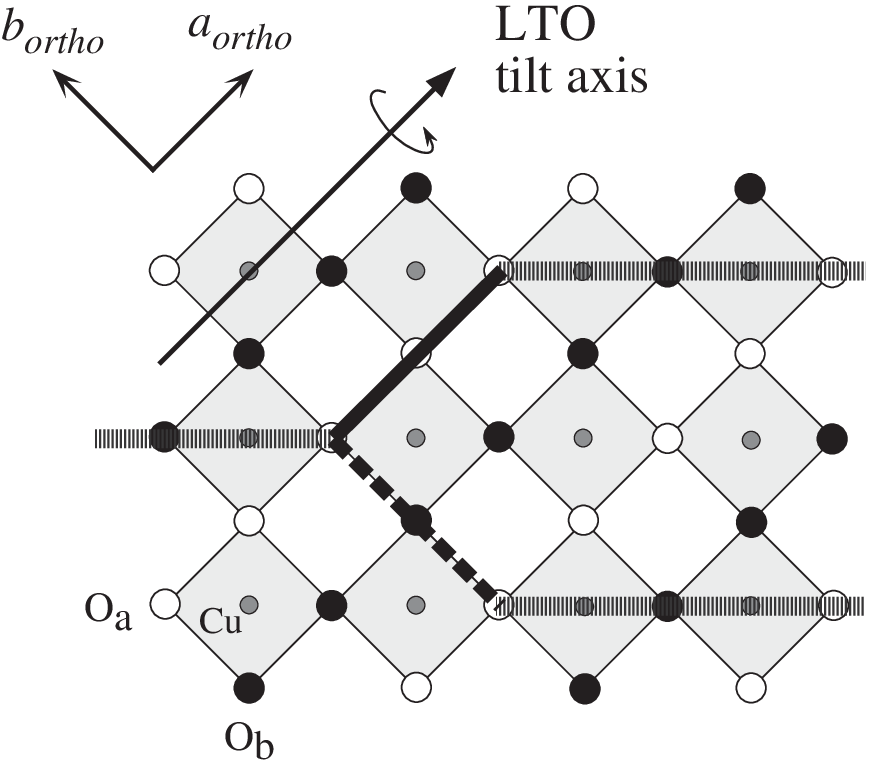}}
\caption
{
Corrugated pattern of the CuO$_{2}$ plane and charge stripes with bond centered (O atoms) step (thick lines) in the LTO phase. Oxygen atoms below (O$_{b}$) and above (O$_{a}$) the plane in a staggered pattern create an anisotropy for the step of charge stripe along the orthorhombic {\it a}-axis (O$_{a}$-O$_{a}$-O$_{a}$ shown by solid line) and {\it b}-axis (O$_{a}$-O$_{b}$-O$_{a}$ shown by dashed line). Situation is similar for stripes with site centered (Cu atoms) step. 
}
\end{figure}
\noindent
These results lead to the important consequence that CDW/SDW orders are stabilized as the wavevectors turn toward the high-symmetry axis. 

In the framework of the stripe model, steps or kinks of the charge stripe, which would be caused by Coulomb interactions between holes in an antiferromagnetic background on the underlying CuO$_{2}$ plane, is the origin of {\it Y}-shift.\cite{Bosch01,Eskes} 
(For example, a tilt with $\theta$$_{Ym}$ of 3.0 $^{\circ}$ corresponds to one step every $\sim$ 19 Cu site on the stripe.) 
The LTT lattice potential forces the charge stripe to be straight, stabilizing the CDW and SDW orders. 
The anisotropic tilt may possibly be characterized by orthorhombic distortion, as the step directions of the charge stripe are no longer equivalent on the corrugated CuO$_{2}$ plane (illustrated in Fig. 7 for the LTO phase). 
Assuming a forward step along the orthorhombic-axis, experimental results suggest a step along {\it a}-{\it axis}, which is parallel to the ridge or valley of the corrugation. 
Although the character of the shift can be well explained by the stripe order, the {\it Y}-shift of SDW peaks with no well-defined CDW order in the LTO phase~\cite{Kimura00,Fujita_cond-mat} needs to be elucidated. 
Alternatively, the anisotropic peak-shift of the SDW peak can also be reproduced in term of fermiology, even in a homogeneous charge distribution if the structural anisotropy of the orthorhombic phase is taken into account.~\cite{Yamase00} 
Since the anisotropy of the next-nearest-neighbor hopping integral ({\it t}$^{\prime}$) between Cu atoms (parallel to orthorhombic axes) causes the {\it Y}-shift, this shift is not expected to occur in the tetragonal phase. 
Then if additional effects such as reduction of $\verb;|;${\it t}$^{\prime}$$\verb;|;$\cite{Tohyama,Normand}, Ising spin interaction\cite{Riera}, electron-phonon coupling\cite{Neto} and lattice potential with zigzag symmetry\cite{Hasselmann} are present in the LTT and LTLO phases, the striped phase is realized or enhanced, meaning a transition or a crossover between homogeneous and inhomogeneous charge distribution. 
In order to clarify the origin of the {\it Y}-shift as well as the charge distribution, further experiments are required such as observation of the {\it Y}-shift in a system with flat CuO$_{2}$ and/or in low-energy magnetic excitation spectra. 

In conclusion, we found the anisotropic tilt of both CDW and SDW wavevectors from the Cu-O bond direction in the LTLO phase, while they were aligned along the high-symmetry axis in the LTT phase. 
The coincident tilt direction of the two wavevectors demonstrates the close relation between CDW and SDW orders. 
Furthermore, the tilt angle of the SDW wavevector changes systematically with the lattice distortion while maintaining a constant total hole concentration, suggesting a structural effect on the deviation. 
These new findings provide challenges to existing high-{\it T}$_{c}$ superconductivity theories. 
We believe that the enhancement of the {\it Y}-shift with degrading or vanishing CDW order in the higher {\it T}$_{c}$ region of the LBSCO system is one of the key points for understanding the distribution or formation of charges doped in Mott insulators.

\section*{Acknowledgements}

We would like to thank H. Kimura for sharing data, and H. Hiraka, Y. S. Lee, G. Shirane, J. M. Tranquada and H. Yamase for stimulating discussions. This work was supported in part by the Japanese Ministry of Education, Culture, Sports, Science and Technology, Grant-in-Aid for Scientific Research on Priority Areas (Novel Quantum Phenomena in Transition Metal Oxides), 12046239, 2000, for Scientific Research (A), 10304026, 2000, for Encouragement of Young Scientists, 13740216, 2001 and for Creative Scientific Research (13NP0201) "Collaboratory on Electron Correlations - Toward a New Research Network between Physics and Chemistry -", by the Japan Science and Technology Corporation, the Core Research for Evolutional Science and Technology Project (CREST).



\begin{references} 
%
\bibitem{Tranquada95} J. M. Tranquada B. J. Sternlieb, J. D. Axe, Y. Nakamura, and S. Uchida, Nature (London) {\bf 375}, 561 (1995).
\bibitem{Emery97} V. J. Emery, S. A. Kivelson, and O. Zachar, Phys. Rev. B {\bf 56}, 6120 (1997). 
\bibitem{Castellani97} C. Castellani, C. Di Castro, and M. Grilli, Z. Phys. B {\bf 103}, 137 (1997). 
\bibitem{Vojta99} M. Vojta, and S. Sachdev, Phys. Rev. Lett. {\bf 83}, 3916 (1999). 
\bibitem{Ichikawa00} N. Ichikawa, S. Uchida, J. M. Tranquada, T. Niem$\ddot{o}$ller, P. M. Gehring, S.-H. Lee, and J. R. Schneider, Phys. Rev. Lett. {\bf 85}, 1738 (2000). 
\bibitem{Tranquada97} J. M. Tranquada, J. D. Axe, N. Ichikawa, A. R. Moodenbaugh, Y. Nakamura, and S. Uchida, Phys. Rev. Lett. {\bf 78}, 338 (1997). 
\bibitem{YoungLee99} Y. S. Lee, R. J. Birgeneau, M. A. Kastner, Y. Endoh, S. Wakimoto, K. Yamada, R. W. Erwin, S.-H. Lee, and G. Shirane, Phys. Rev. B {\bf 60}, 3643 (1999). 
\bibitem{Kimura99} H. Kimura, K. Hirota, H. Matsushita, K. Yamada, Y. Endoh, S.-H. Lee, C. F. Majkrzak, R. Erwin, G. Shirane, M. Greven, Y. S. Lee, M. A. Kastner, and R. J. Birgeneau, Phys. Rev. B {\bf 59}, 6517 (1999). 
\bibitem{Fujita_cond-mat} M. Fujita, H. Goka, K. Yamada, and M. Matsuda, Phys. Rev. Lett. {\bf 88}, 167008 (2002).
\bibitem{Tranquada_private} J. M. Tranquada, private communication. 
\bibitem{Kimura00} H. Kimura, H. Matsushita, K. Hirota, Y. Endoh, K. Yamada, G. Shirane, Y. S. Lee, M. A. Kastner, and R. J. Birgeneau, Phys. Rev. B {\bf 61}, 14366 (2000). 
\bibitem{Bosch01} M. Bosch, W. van Saarloos, and J. Zaanen, Phys. Rev. B {\bf 63}, 92501 (2001). 
\bibitem{Fujita01} M. Fujita, H. Goka, and K. Yamada, Int. J. Mod. Phys. B {\bf 14}, 3466 (2000). 
\bibitem{Goka_unpublish} H. Goka, M. Fujita, Y. Ikeda and K. Yamada, Physica C {\bf 357-360}, 256 (2001). 
\bibitem{Maeno91} Y. Maeno, A. Odagawa, N. Kakehi, T. Suzuki, and T. Fujita, Physica C {\bf 173}, 322 (1991). 
\bibitem{Kimura_unpub} H. Kimura, (unpublished). 
\bibitem{Tranquada96} J. M. Tranquada, J. D. Axe, N. Ichikawa, Y. Nakamura, S. Uchida, and B. Nachumi, Phys. Rev. B {\bf 54}, 7489 (1996). 
\bibitem{Kimura01} H. Kimura, K. Hirota, M. Aoyama, T. Adachi, T. Kawamata, Y. Koike, K. Yamada, and Y. Endoh, J. Phys. Soc. Jpn. {\bf 70}, 52 (2001) Suppl. A. 
\bibitem{Matsushita99} H. Matsushita, H. Kimura, M. Fujita, K. Yamada, K. Hirota, and Y. Endoh, J. Phys. Chem. Solid {\bf 60}, 1071 (1999).
\bibitem{calc} $\theta$$_{ortho}$ for LSCO and LCO is calculated from $\theta$$_{ortho}$={\it tan}$^{-1}$({\it b}$_{ortho}$/{\it a}$_{ortho}$) - 45, where {\it a}$_{ortho}$ and {\it b}$_{ortho}$ are orthorhombic lattice constants ({\it Bmab} notation) reported in references.  
\bibitem{Kimura_private} Recently {\it Y}-shift of SDW peaks has been found in La$_{1.85}$Sr$_{0.15}$Cu$_{0.983}$Zn$_{0.017}$O$_{4}$ by H. Kimura. (unpublished)
\bibitem{Eskes} H. Eskes, R. Grimberg, W. van Saarloos, and J. Zaanen, Phys. Rev. B {\bf 54}, R724 (1996).
\bibitem{Yamase00} H. Yamase, and H. Kohno, J. Phys. Soc. Jpn. {\bf 69}, 332 (2000). 
\bibitem{Tohyama} T. Tohyama, C. Gazza, C. T. Shih, Y. C. Chen, T. K. Lee, S. Maekawa, and E. Dagatto, Phys. Rev. B {\bf 59}, R11649 (1999). 
\bibitem{Normand} B. Normand, and A. P. Kampf, Phys. Rev. B {\bf 65}, 020509 (2002). 
\bibitem{Riera} J. A. Riera, Phys. Rev. B {\bf 64}, 104520 (2001).
\bibitem{Neto} A. H. Castro Neto, Phys. Rev. B {\bf 64}, 104509 (2001).
\bibitem{Hasselmann} N. Hasselmann, A. H. Castro Neto, C. Morais Smith, Phys. Rev. B {\bf 65}, 220511 (2002). 
%
\end{references}
\end{document}